\documentclass[twocolumn,final,aps,prb,showpacs,superscriptaddress,amsmath,amssymb,amsfonts,floatfix]{revtex4-2}
\usepackage{epsfig}
\usepackage[dvipsnames,usenames]{color}
\usepackage{xcolor}
\usepackage{amsmath}
\usepackage{comment}
\usepackage{array}
\usepackage{multirow}
\usepackage{tabularx}
\usepackage{float}
\usepackage[colorlinks=true,bookmarks=true, pdfborder={0 0 0},linkcolor=blue,urlcolor=blue,     breaklinks=true,bookmarksnumbered=true]{hyperref}

\begin{document}
\title{Enhanced superconductivity via layer differentiation in trilayer Hubbard model}
\author{Xun Liu}
\affiliation{School of Physical Science and Technology, Soochow University, Suzhou 215006, China}

\author{Mi Jiang}
%\email{jiangmi@suda.edu.cn}
\affiliation{School of Physical Science and Technology, Soochow University, Suzhou 215006, China}
\affiliation{Jiangsu Key Laboratory of Frontier Material Physics and Devices, Soochow University, Suzhou 215006, China}

\begin{abstract}
Motivated by the highest superconducting transition temperature ($T_c$) in multilayer cuprates, %HgBa$_2$Ca$_2$Cu$_3$O$_{8+\delta}$ (Hg-1223) 
%consisting of three CuO$_2$ layers,
we investigated the trilayer Hubbard model by adopting the large-scale dynamical cluster quantum Monte Carlo simulations. Focusing on the systems with hole dopings within the two outer layers (OL) higher than the inner layer (IL), which is believed to be relevant to the realistic multilayer cuprates, our exploration discovered that the IL and OL manifest strong differentiation in a wide range of hole doping combinations. Specifically, the OLs remain metallic while the IL shows a distinct transition from the pseudogap to superconducting state. More importantly, the highest $T_c$ of the composite trilayer system can be largely enhanced compared to the single layer model and the imbalanced hole dopings between IL and OL are generically beneficial for global SC. We further provide strong numerical evidence on the possibility of $d$-wave superconductivity solely hosted in the IL. 
%with the cooperation from OLs. 
Our investigation provides new insight into the origin of highest $T_c$ in multilayer cuprates.

%We further provide strong numerical evidence on the picture that the IL itself can drive the $d$-wave superconductivity while the OLs only serve as the charge reservoir.Our investigation provides new insight into the origin of highest $T_c$ in multilayer cuprates.

%composite picture and/or proximity effects involving the acting role of all layers are not necessary in the present simplest trilayer model.
\end{abstract}

\maketitle

\section{introduction}
Within the family of cuprate superconductors, the HgBa$_2$Ca$_2$Cu$_3$O$_{8+\delta}$ (Hg-1223) composed of three non-equivalent CuO$_2$ layers (i.e., one inner layer (IL) and two outer layers (OL) exhibits the highest superconducting critical temperature $T_c\approx134$ K at ambient pressure \cite{HgBaCaCuO}. It has been found that the superconducting (SC) gap on the underdoped IL is larger than the counterpart on the overdoped OL \cite{Hg1223_Arxiv}. In contrary, the Bi$_2$Sr$_2$Ca$_2$Cu$_3$O$_{10+\delta}$ (Bi-2223) with $T_c\approx110$ K~\cite{ZhouXJ,SZX_Bi} was found to have its SC gap on the OL less than that in Hg-1223 \cite{Hg1223_Arxiv}. Notably, the $T_c$ in Bi-2223 is nearly constant in the optimally and overdoped regions~\cite{BiSrCaCuO}. Interestingly, this trend is opposite to the recently discovered Ruddlesden-Popper (RP) nickelates, where the $T_c$ of trilayer compound is much lower than its bilayer counterpart under high pressure \cite{WM327,ZJ4310}. In addition, In the (Cu,C)Ba$_2$Ca$_{n-1}$Cu$_n$O$_{2n+4-y}$ family, the four-layer CuC-1234 ($n=4$) has the higher $T_c\approx117$ K than the $T_c\approx71$ K in the trilayer CuC-1223 ($n=3$) \cite{CuC-1234,pengrui} and the (Cu$_{1-x}$Tl$_x$)Ba$_2$Ca$_3$Cu$_4$O$_{12-y}$ (Cu$_{1-x}$Tl$_x$-1234) has highest $T_c\approx126$ K \cite{CuTl-1234}. 

The microscopic mechanism of the highest $T_c$ in trilayer cuprates is unclear up to now. To resolve this long-standing puzzle, the composite picture has been proposed~\cite{Kivelson, Kivelson2, Maier2008,Dror,Maier2022}, where the large pairing scale is derived from the underdoped layer while the large phase stiffness arises mainly from the optimally or overdoped component, e.g. the previous work suggests that the interaction between the IL and OL is necessary to expalin the hightes $T_c$ in Bi-2223 \cite{Fujimori}. Experimentally it has been suggested that the difference of density distribution between the IL and OL plays an important role to reach the maximum $T_c$ \cite{density1}, although another study proposed that $T_c$ could be enhanced if all layers have balanced optimal dopings~\cite{imbalance}. Nonetheless, there is another insightful thought claiming that the origin of high $T_c$ is due to the large SC gap in the IL by the protection from the disorder from the OL and thereby the SC gap in the OL comes from the proximity effect \cite{IP_dominant}. Meanwhile, the interlayer hybridization is essential for the enhancement of the pairing scale in the OLs from the proximity effect \cite{Maier2008}, e.g. the nearest interlayer hopping between the IL and OL is more significant than that interlayer hopping between the OLs \cite{ZhouXJ,Bi2223_Optimally} in Bi-2223. Interestingly, a recent theoretical study attributed the growth of $T_c$ to the reduction of the charge transfer gap so that promoted superexchange~\cite{Abinitio}. %\lx{Interestingly, the previous numerical work on the two-component (superconductor/metal) model has show  that there is enhancing $T_c$ with repulsive Coulomb interaction \cite{Maier2022} but not with attractive one \cite{Dror}}.

To compare the roles played by the IL and OL, we concentrated on the single-orbital trilayer Hubbard model to investigate the possible origin of SC by tuning the hole concentration distribution between the IL and OL. We uncovered that in the most realistic systems with hole dopings within OL higher than that of IL, the IL itself can host the $d$-wave superconductivity, which is closely connected to its pseudogap behavior absent in the OLs and qualitatively consistent with the recent experiment~\cite{pengrui}. The strong layer differentiation in the trilayer Hubbard model thus provides new constraints on the scenarios of either composite picture or proximity effects.  

%The paper is organized as follows. Section II illustrates the trilayer model and the many-body method we used; Sec. III presents the results of the layer differentiation via the spectral properties and the density-modulated SC; Sec. IV includes a summary and outlook. 

\begin{figure}
\psfig{figure=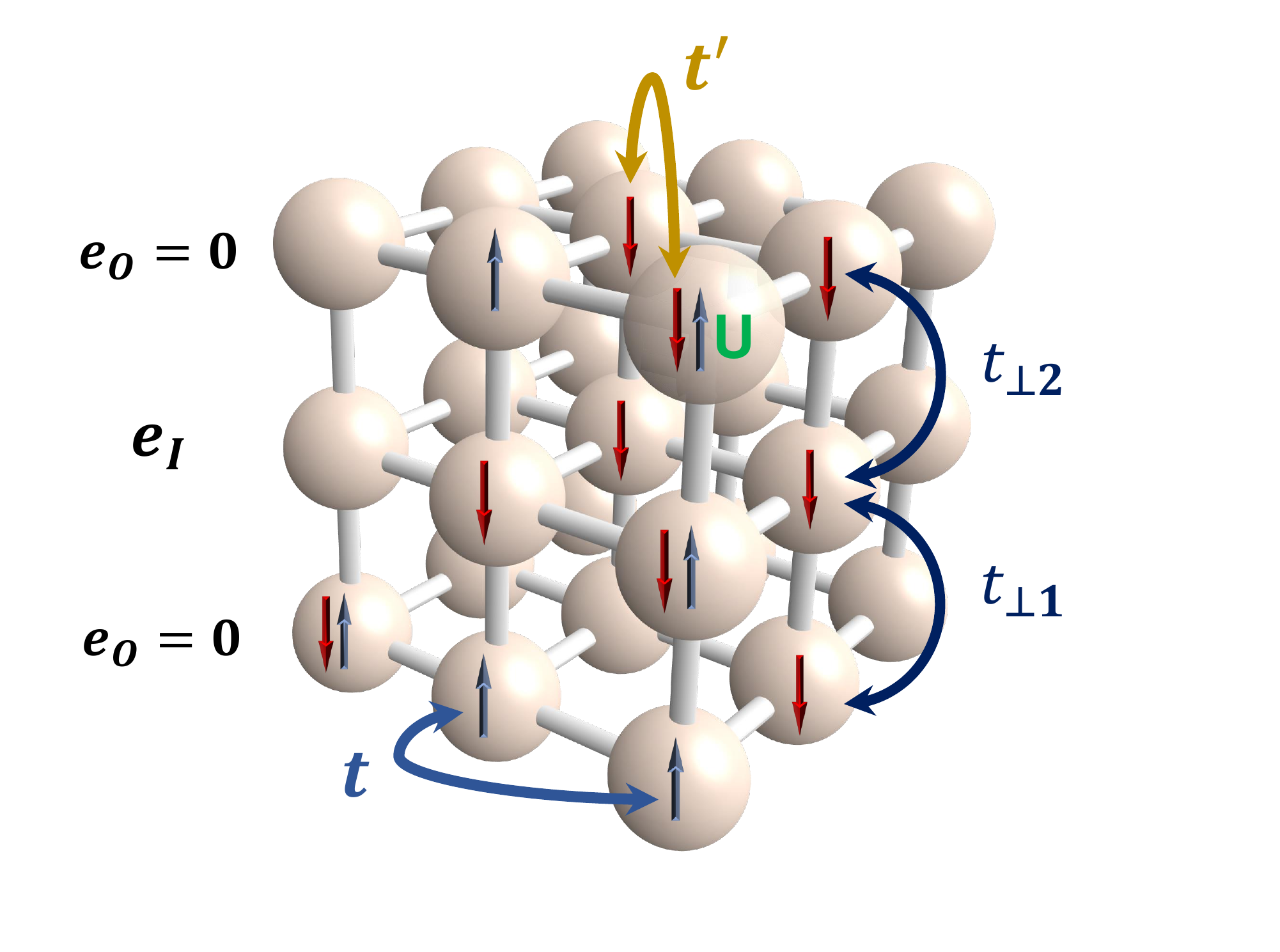,height=6.2cm,width=0.52\textwidth,clip}
\caption{Schematic diagram of the trilayer Hubbard model labeled with various hoppings and on-site interaction $U$. The site energy $e_I \neq0$ controls the density distribution.}
\label{lattice}
\end{figure}

\begin{figure*}[t]
\psfig{figure=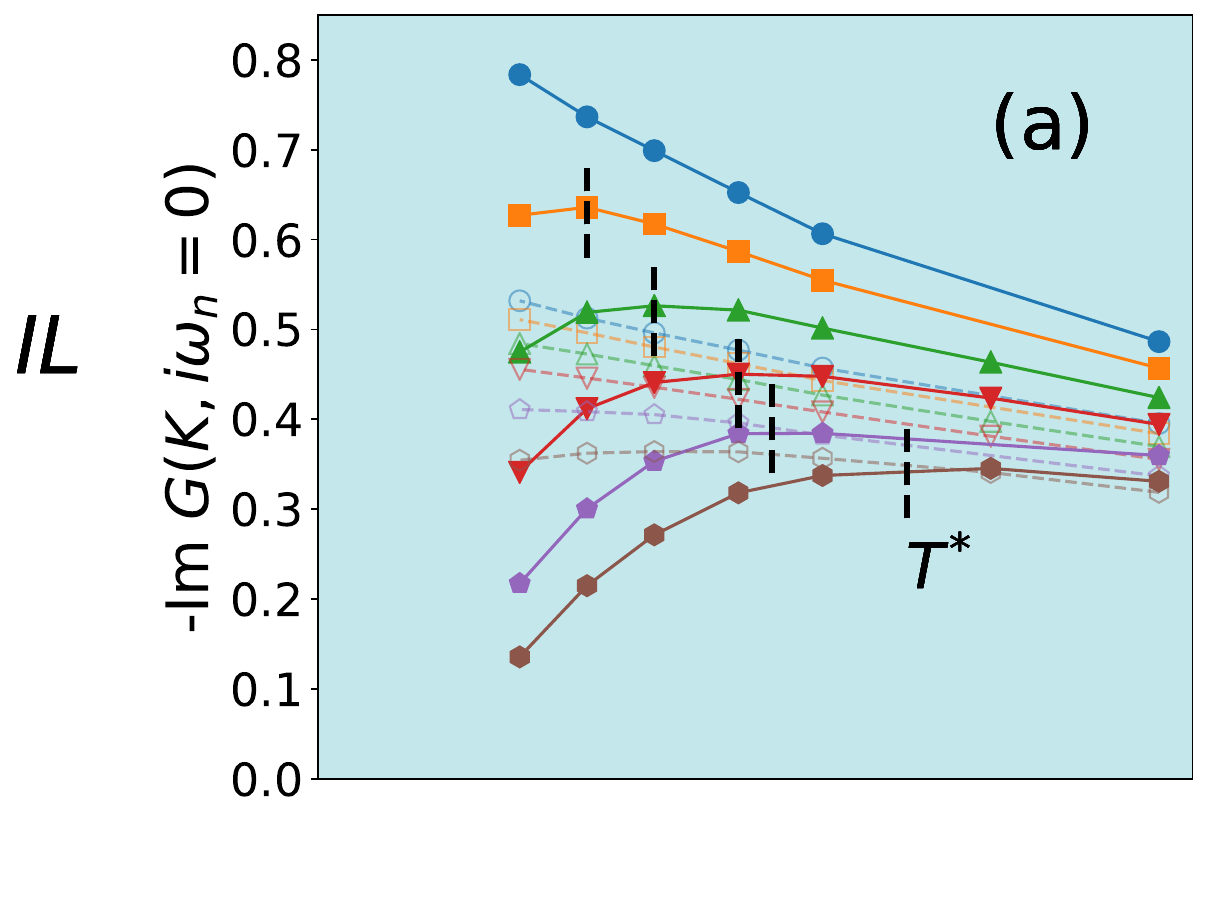,height=3.8cm,width=0.33\textwidth,clip}
\vspace{-1.5em}
\hspace{0.1em}
\psfig{figure=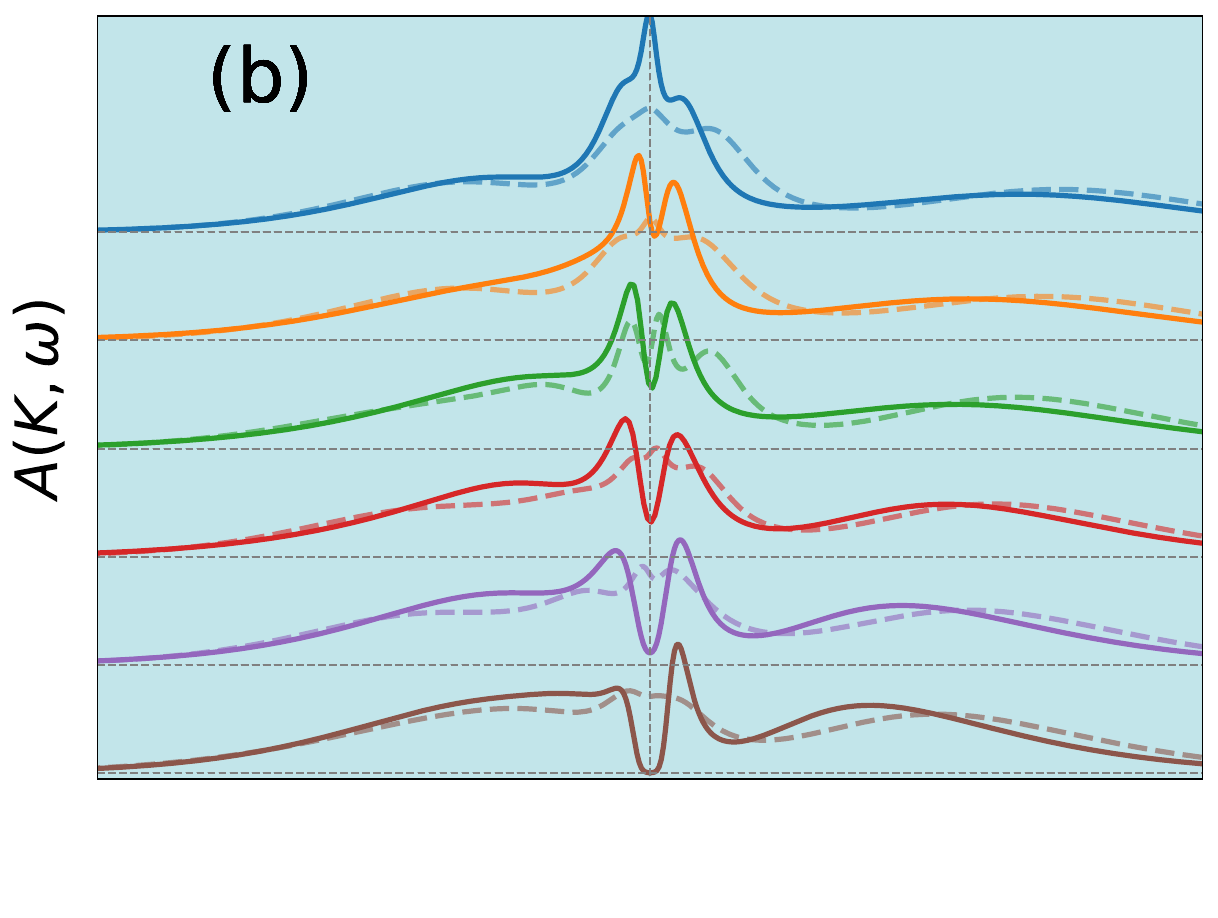,height=3.8cm,width=0.3\textwidth,clip}
\hspace{0.1em}
\psfig{figure=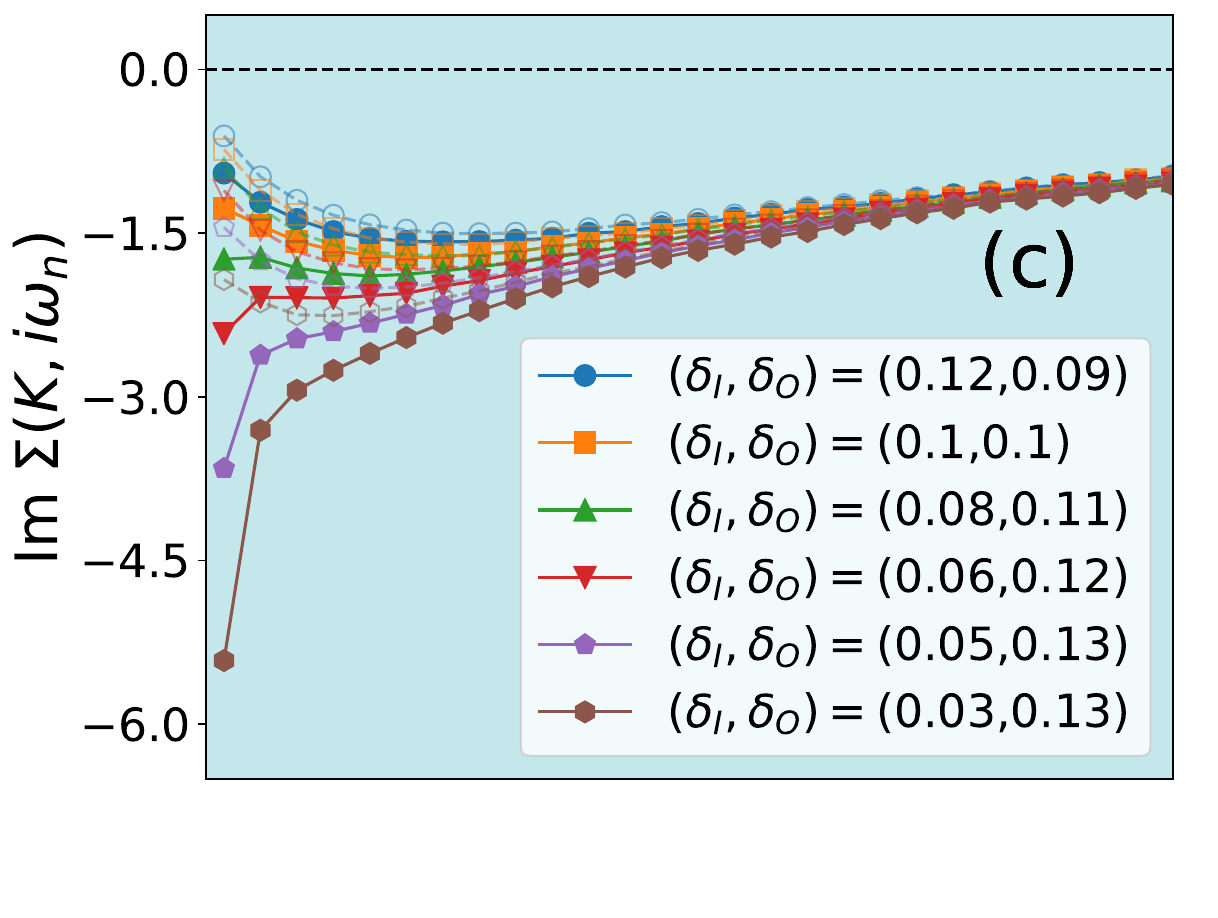,height=3.8cm,width=0.3\textwidth,clip}
\psfig{figure=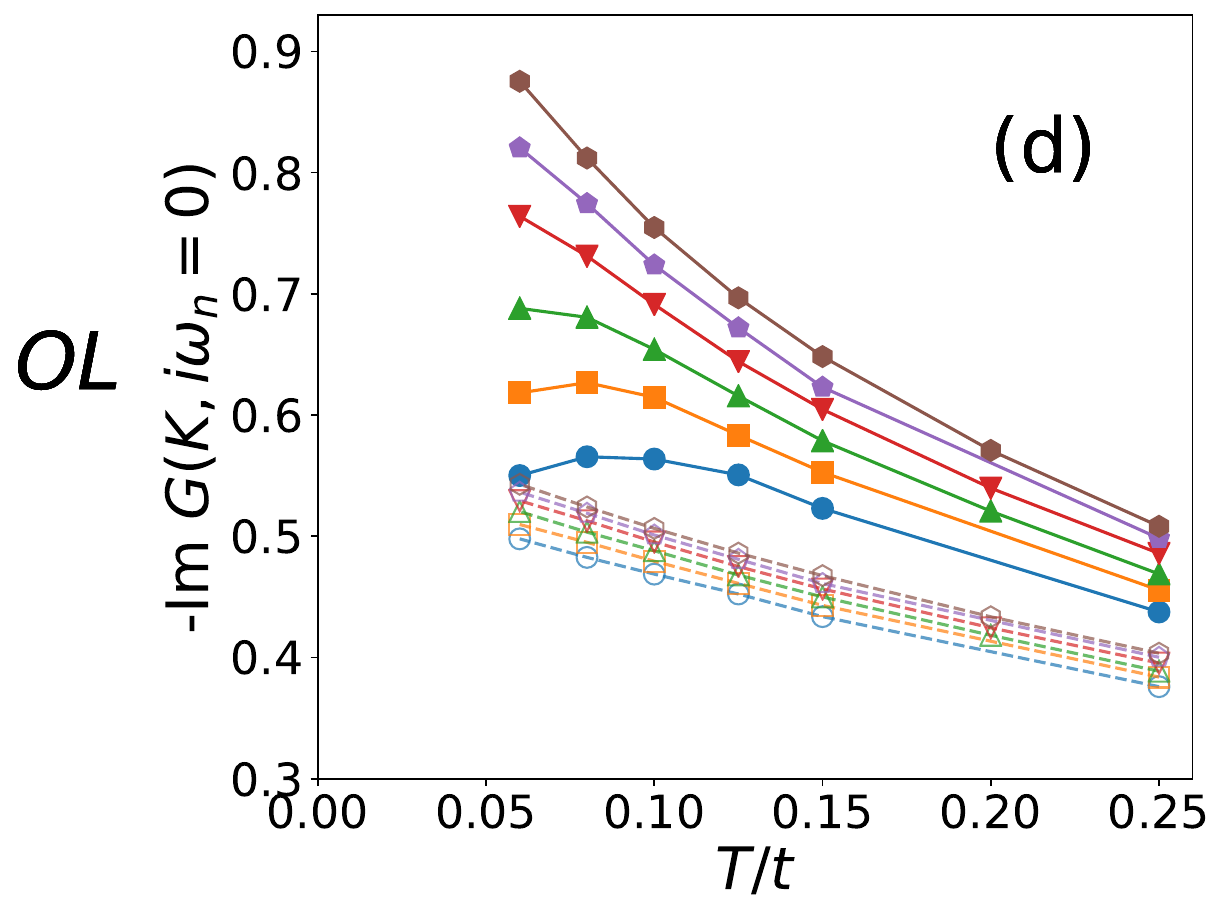,height=3.8cm,width=0.33\textwidth,clip}
\hspace{0.1em}
\psfig{figure=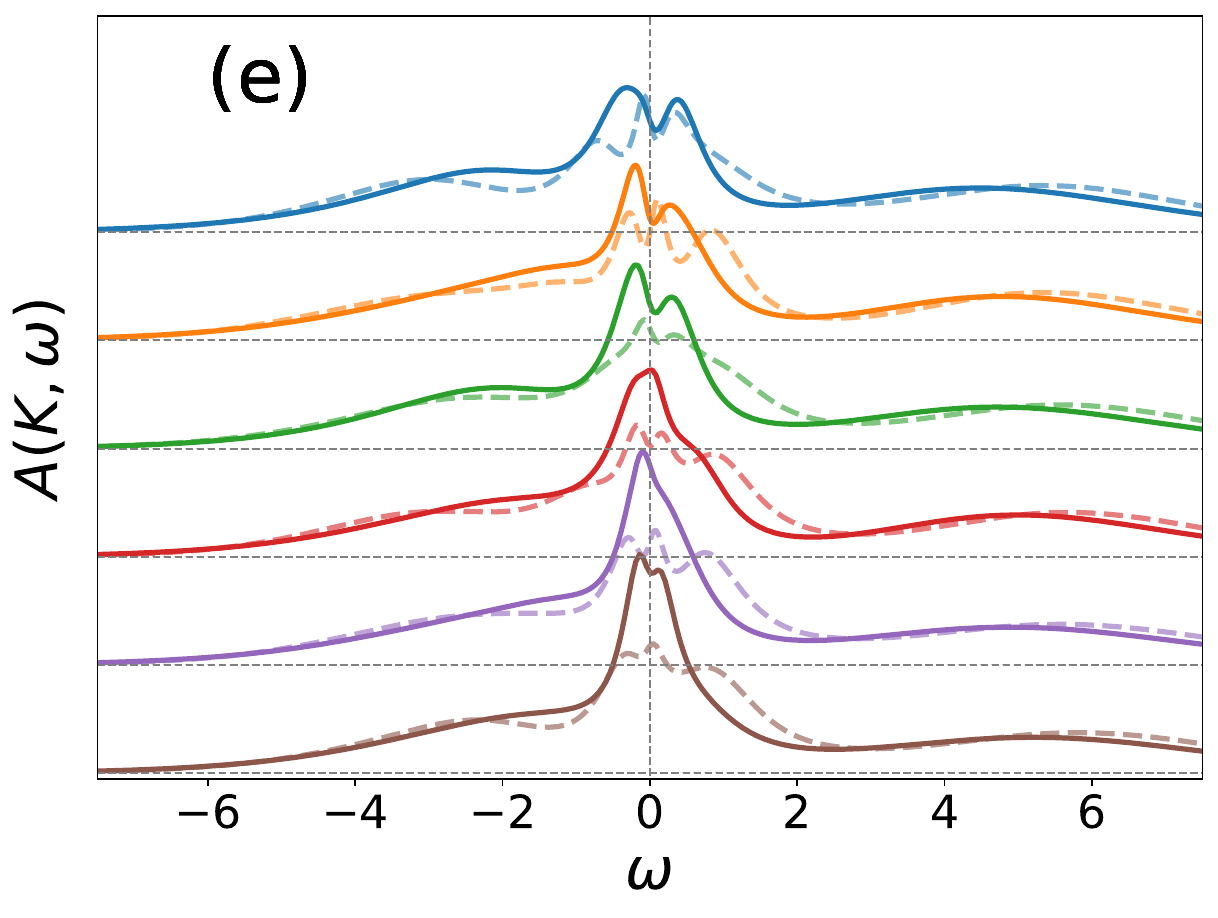,height=3.8cm,width=0.3\textwidth,clip}
\hspace{0.1em}
\psfig{figure=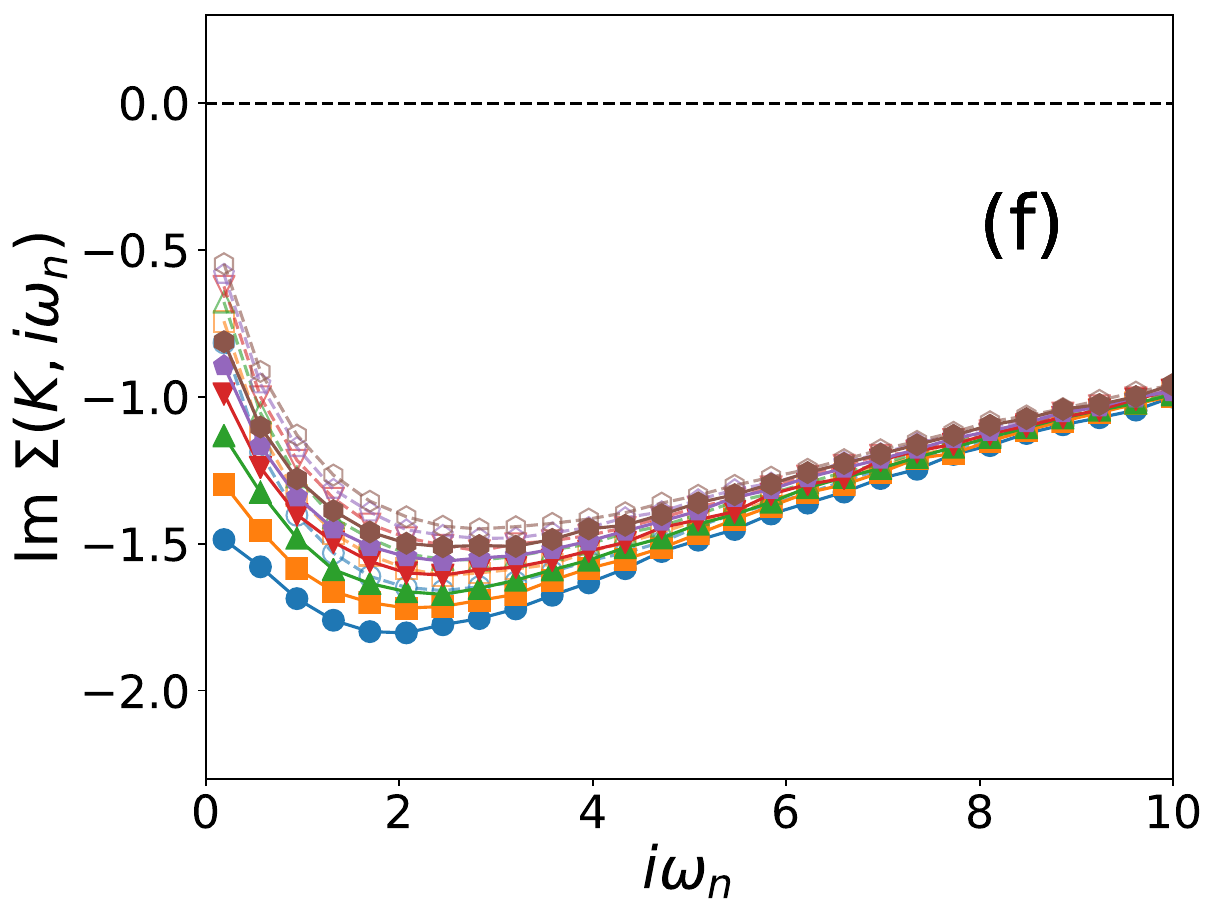,height=3.8cm,width=0.3\textwidth,clip}
\caption{Layer and momentum differentiation between IL (upper) and OL (lower) at $\delta_{avg}=0.1$. (a) and (d): Temperature evolution of the extrapolated imaginary zero-frequency -Im$G(\textbf{K}, i\omega_n=0)$ obtained from a linear extrapolation of the first two Matsubara frequencies; (b) and (e): The spectral function $A(\textbf{K},\omega)$ at $T/t=0.06$; (c) and (f): The imaginary part of self-energy Im$\Sigma(\textbf{K},i\omega_n)$ at $T/t=0.06$. The solid and dashed line or open symbols indicate the anti-nodal $\textbf{K}=(\pi,0)$ and nodal $\textbf{K}=(\frac{\pi}{2},\frac{\pi}{2})$ directions respectively.}
\label{2layers}
\end{figure*}

\section{model and method}

%\subsection{Trilayer Hubbard model}
We consider the trilayer Hubbard model on two-dimensional square lattice as shown in Fig.~\ref{lattice}, the Hamiltonian is
\begin{align}\label{H}
\hat{H}= & -\sum_{ij\sigma m}t_{ij}c^\dagger_{i\sigma m}c_{j\sigma m} -\sum_{i\sigma, m=1,3}t_\perp c^\dagger_{i\sigma 2}c_{i\sigma m}+\mathrm{H.c.} \notag\\
         & +U\sum_{i,m}n_{im\uparrow}n_{im\downarrow}+e_I\sum_{i\sigma }n_{i\sigma 2}-\mu\sum_{i\sigma m}n_{i\sigma m} \notag
\end{align}
where $c^\dagger_{i\sigma m}/c_{i\sigma m}$ creates/annihilates an electron with spin  $\sigma$(=$\uparrow,\downarrow$) at the $i$ site on the $m$th layer ($m=1,2,3$). The intralayer hopping $t_{ij}$ includes the nearest neighbor $t$ and the next-nearest-neighbor hopping $t'$. The desired average electron density of the whole system $n$ can be achieved by tuning the chemical potential $\mu$. Meanwhile, to incorporate the differentiation between the IL and OL, the site energy $e_I$ is used to adjust the density distribution. The interlayer hopping and the on-site Coulomb interaction are denoted by $t_{\perp1}=t_{\perp2}=t_{\perp}$ and $U$ respectively. Throughout this work, we set the nearest neighbor hopping $t=1$ as the energy unit and adopt the next-nearest-neighbor hopping $t'=-0.15t$ as well as $U=7t$ \cite{WuWei2018, Aw}. Additonally, we choose the interlayer hopping $t_{\perp}=0.1t$ as the representative value to mimic the trilayer cuprates \cite{ZhouXJ,WenHH,tperp1,tperp2}. For simplicity, we neglect the direct hopping between the two OLs~\cite{ZhouXJ}. 

%\subsection{Dynamical cluster approximation (DCA)}
DCA with the continuous-time auxilary-field (CT-AUX) quantum Monte Carlo (QMC) cluster solver~\cite{Hettler98,Maier05,code,GullCTAUX} is employed to numerically solve the trilayer model.
%As a celebrated quantum many-body numerical method, 
Precisely, our calculations were conducted with $N_c=24 (=8\times3)$ cluster to accomplish the momentum space resolution including both nodal $\mathbf{K}= (\pi/2,\pi/2)$ and antinodal $\mathbf{K} = (\pi,0)$ directions. We remark that the larger cluster is much more difficult because of the lowest accessible temperature will be limited by the severe QMC sign problem. 
More detailed explanation on the methodology is given in the Appendix.

It has been widely accepted that the $d$-wave pairing plays a dominant role in the cuprate superconductors and closely relevant single-orbital Hubbard model~\cite{scalapino2007numerical,Hubreview1}. Likewise, the superconducting gap of the IL and OL in the trilayer model hosts the $d$-wave pairing symmertry in the previous numerical work \cite{tperp2}. Thus, here we focus on the $d$-wave pairing symmetry.

%\section{results}
We illustrate our findings via two characteristic average hole doping levels $(\delta_{avg} = 0.1, 0.15)$, each of which has multiple density distributions tuned by the parameter $e_I$. The hole dopings of IL and OL as well as the average doping are labeled as $\delta_I$, $\delta_O$, and $\delta_{avg}$ respectively. In the next two sections, we will not only demonstrate the difference between the OL and IL, but also show the SC solely hosted in the IL can support even higher $T_c$ than the single-layer model.

\section{Layer differentiation}

\begin{figure*}[t!]
\psfig{figure=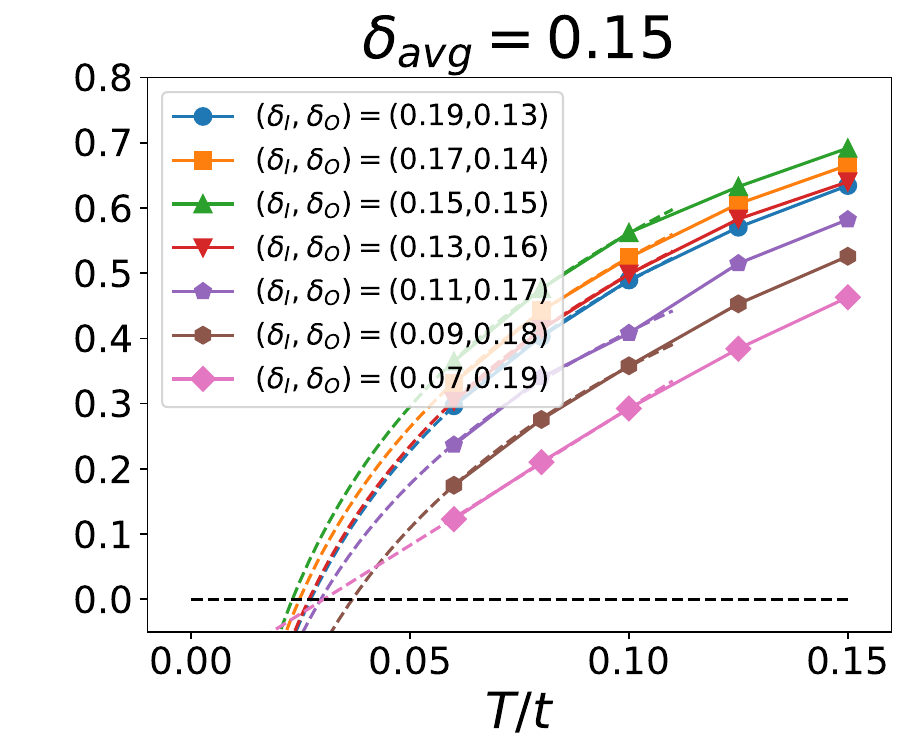,height=4.5cm,width=0.32\textwidth,clip}
\psfig{figure=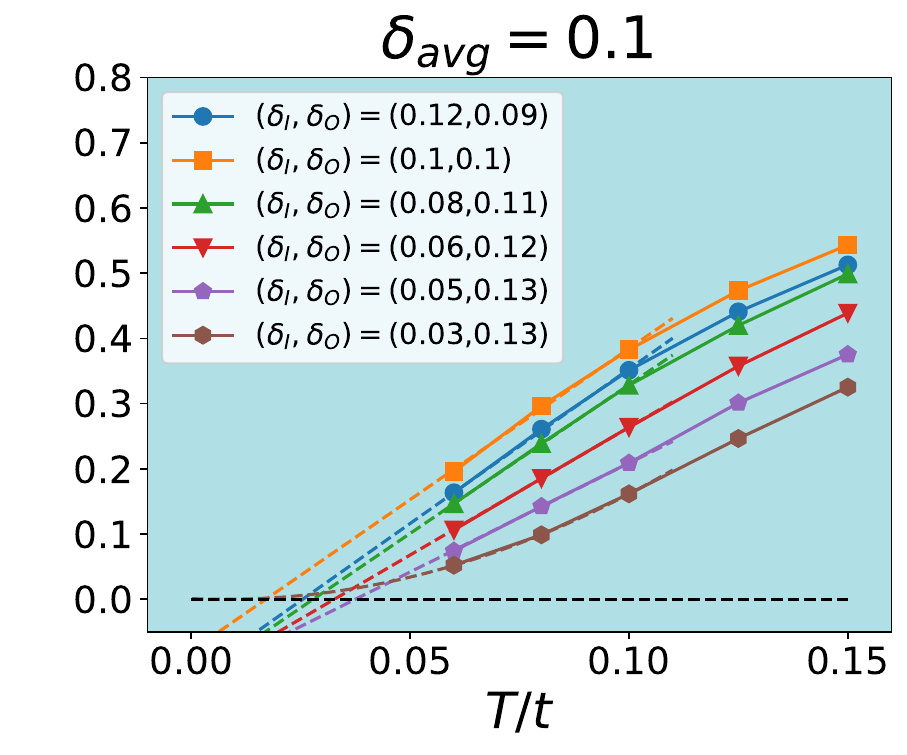,height=4.5cm,width=0.32\textwidth,clip}
\psfig{figure=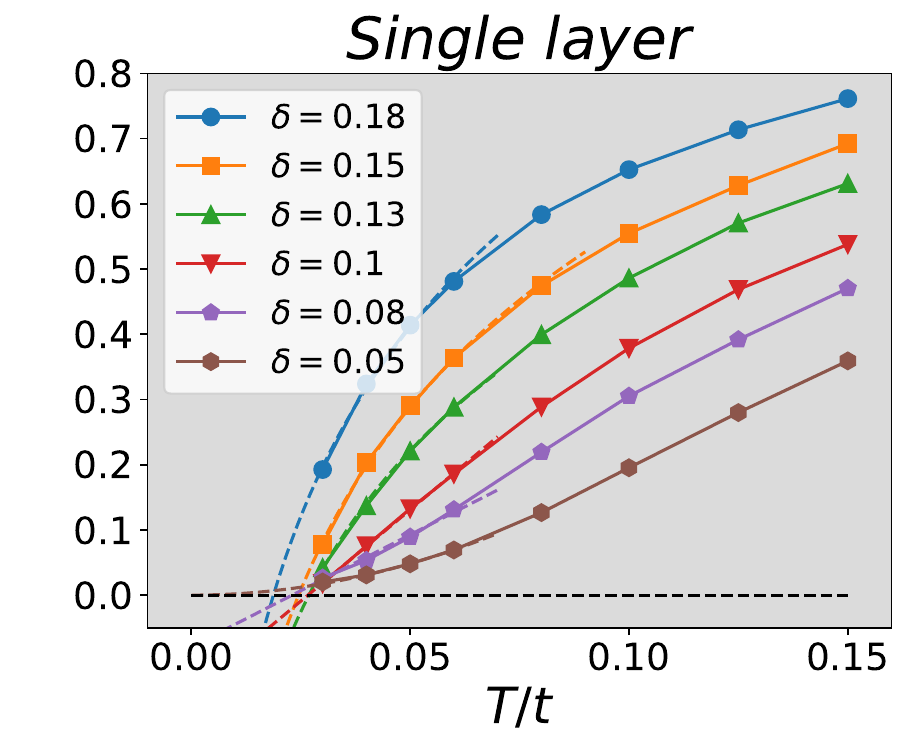,height=4.5cm,width=0.32\textwidth,clip}
\caption{Temperature evolution of $1-\lambda_d$ for various hole density distribution ($\delta_I,\delta_O$) at the fixed $\delta_{avg}=0.15, 0.1$ of trilayer model. The single layer case is shown as comparison.}
\label{eigenvalue}
\end{figure*}

To demonstrate the differentiation between the IL and OL, we employed the imaginary part of the single-particle Green function Im$G(\textbf{K},i\omega_n$) and self-energy Im$\Sigma(\textbf{K},i\omega_n)$ as well as the $\textbf{K}$-resolved spectrum $A(\textbf{K},\omega$) calculated from analytical continuation of $G(\textbf{K},i\omega_n)$.  Without loss of generality, we only illustrate the results for $\delta_{avg}=0.1$ as an example since $\delta_{avg}=0.15$ qualitatively manifests the same properties.

Fig.~\ref{2layers} summarizes the mismatch between IL (upper) and OL (lower) induced by the hole density distribution.
The comparison between panels (a) and (d) indicates that, in the regime of $\delta_I<\delta_O$, the IL manifests clear PG behavior, namely the anti-nodal $\mathbf{K}=(\pi,0)$ curve with lowering $T$ shows a peak while that of the nodal $\mathbf{K}=(\frac{\pi}{2},\frac{\pi}{2})$ direction remains monotonic. In contrast, the OL in panel (d) always displays the metallic feature until entering into a qualitatively different regime of $\delta_I \ge \delta_O$, where the IL and OL exchange their PG and metallic behavior. 
%We remark that the appearance of PG feature also depends on the exact hole doping $\delta_I$, which was found to be roughly $\delta_I(O) \le 0.1$ for our parameters.
It is obvious that the PG behavior becomes stronger when the doping level decreases from 0.1 to 0.03 on the IL.
In panel (a), we choose the curve's peak as the onset temperature $T^*$ of the PG phase \cite{WuWei2018} and thereby the $T^*$ diminishes with the further hole doping until around $\delta_I\sim 0.1$. Note that, regardless of whether it is the IL or OL, the turning doping level signaling the onset of PG behavior is roughly $\delta\sim 0.1$. 

Using the proxy of the extrapolated Im$G(\textbf{K},i\omega_n=0$) is an approximation in essence. Therefore, Fig.~\ref{2layers}(b) and (e) provide more concrete evidence via the momentum resolved spectral functions, which illustrates clear distinction between the PG and metallic peak at $\omega=0$.

The layer and momentum differentiation can be further supported by the self-energy as shown in Fig.~\ref{2layers}(c) and (f). The convergence of Im$\Sigma(\mathbf{K},i\omega_n)$ to a finite value as $i\omega_n \to 0$ is a typical non-Fermi liquid behavior; while the diverging or decaying behavior indicate the gap opening originating from strong correlation or gap closing separately~\cite{self-energy} at the particular $\mathbf{K}$ point. On the IL, the different evolution of the anti-nodal (solid) and nodal (open) directions with the hole doping combination clearly support the PG behavior, which gradually disappears in the regime of $\delta_I\ge \delta_O$. Besides, the comparison of the scales reveals that, when $\delta_I<\delta_O$, the interaction strength of the OL is weaker than the IL, indicating deviated renormalization of the interaction strength.
%and no diverging behavior, because the more hole doping can suppress the non-Fermi liquid behavior. 

\section{Doping imbalance enhanced SC}

The hole density distribution obviously affects our most interested $d$-wave SC properties. To investigate the possibility of enhanced SC with an ``optimal'' ($\delta_I,\delta_O$) combination, we first explored the temperature evolution of the leading $d$-wave BSE eigenvalue $\lambda_{d}(T)$ for varied density distribution with two typical fixed $\delta_{avg}=0.1, 0.15$, as illustrated in Fig.~\ref{eigenvalue} with panel (c) showing the single layer case as the reference.

\begin{figure}
\psfig{figure=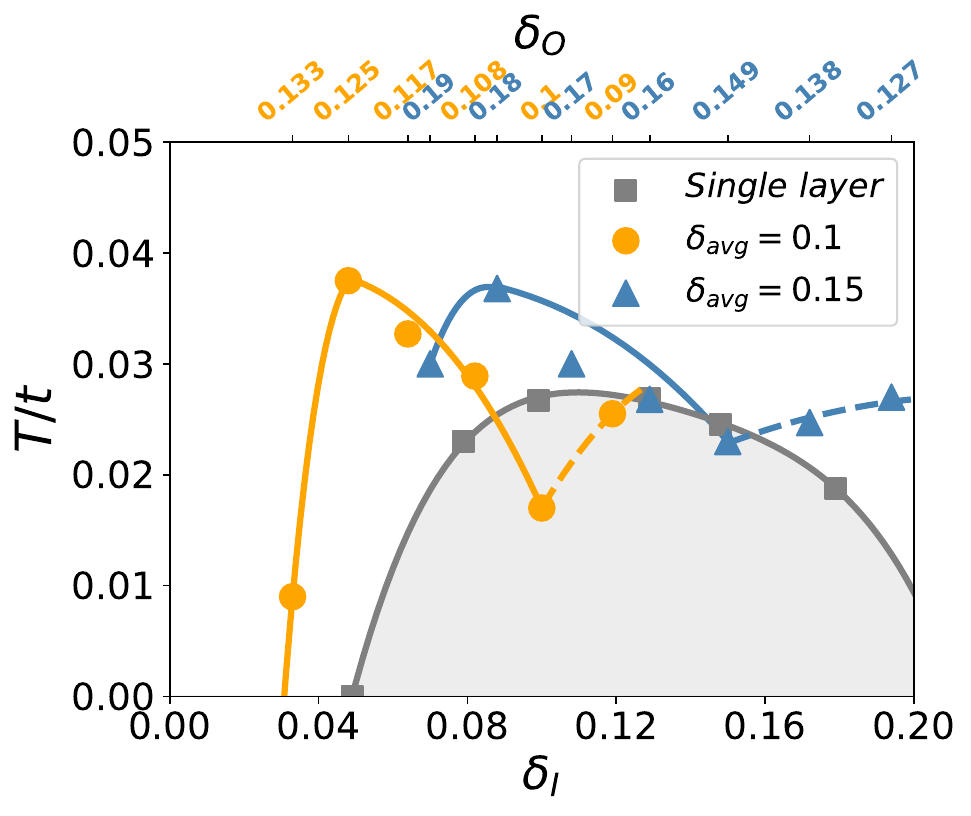,height=7cm,width=.48\textwidth, clip}
\caption{Evolution of $T_c$ on the hole doping $\delta_I$ of IL for two fixed $\delta_{avg}=0.1, 0.15$. The more realistic density distribution $\delta_I<\delta_O$ is indicated by solid line; while the dashed line  denotes the $\delta_I>\delta_O$ regime. The $T_c$ of single layer model is shown as comparison.}
\label{Tc}
\end{figure}

The weakened $d$-wave pairing tendency is manifested by the departure of $1-\lambda_{d}(T)$ curve from zero, which is reflected by the up shift of the curve with increasing $\delta_I$ until $\delta_I=\delta_O$ (orange curves for both $\delta_{avg}=0.1, 0.15$). As discussed later, $\delta_I=\delta_O$ is a critical point where the outer layer starts to play a role in the SC.

The temperature evolution in Fig.~\ref{eigenvalue} can be categorized into two classes. Specifically, the behavior exhibiting the BCS logarithmic temperature evolution suggests the dominance of BCS Cooper pair fluctuations, which is reminiscent of that observed in the overdoped regime of the single layer model~\cite{Maier2019}. In contrast, the low temperature linear or even signals of exponential behavior implies the non-BCS type pair fluctuations as discussed for the pseudogap regime~\cite{Maier2019}. 
For both cases of $\delta_{avg}=0.1, 0.15$, there is a gradual transition from linear or exponential to logarithmic evolution with increasing $\delta_I$, which reflects that the similar behavior of single layer model persists in the trilayer model considering that the interlayer $t_{\perp}=0.1t$ is small.

\begin{figure}
\psfig{figure=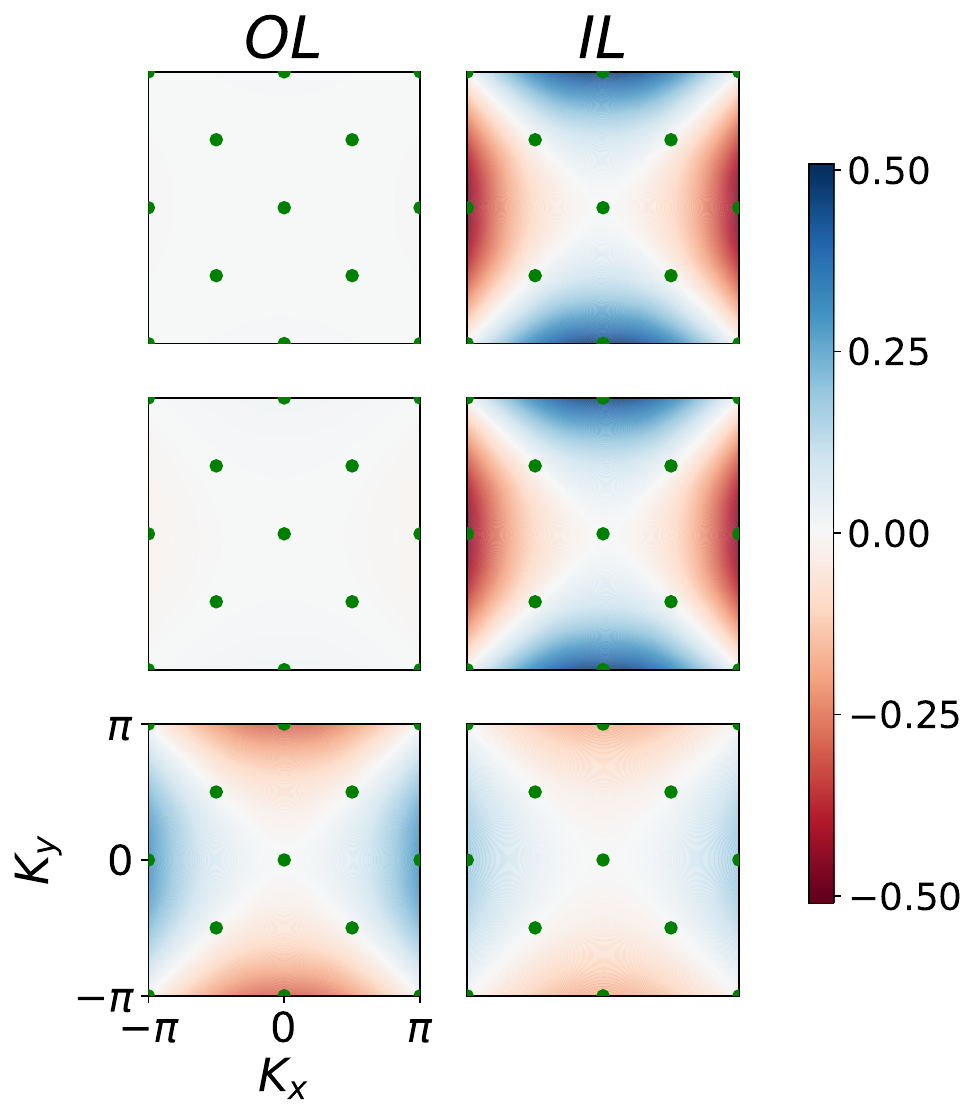,height=9cm,width=0.48\textwidth,clip}
\caption{The leading $d$-wave eigenvectors $\Phi_d(\mathbf{K},\pi T)$ of the OL (left) and IL (right) at $T/t=0.06$ and $\delta_{avg}=0.1$. The dots denote the finite $\mathbf{K}$ points sampling the Brillouin zone via our DCA simulation. The density distributions are $(\delta_I,\delta_O)=(0.06,0.12),(0.08,0.11)$, and $(0.1,0.1)$ in upper, middle, and lower rows respectively, which corresponds to the decreasing $T_c$ segment (orange solid line) in Fig.~\ref{Tc}. When $\delta_I<\delta_O$, the IL is fully responsible for the $d$-wave SC.}
\label{eigenvector}
\end{figure}

\begin{figure*}[t]
\psfig{figure=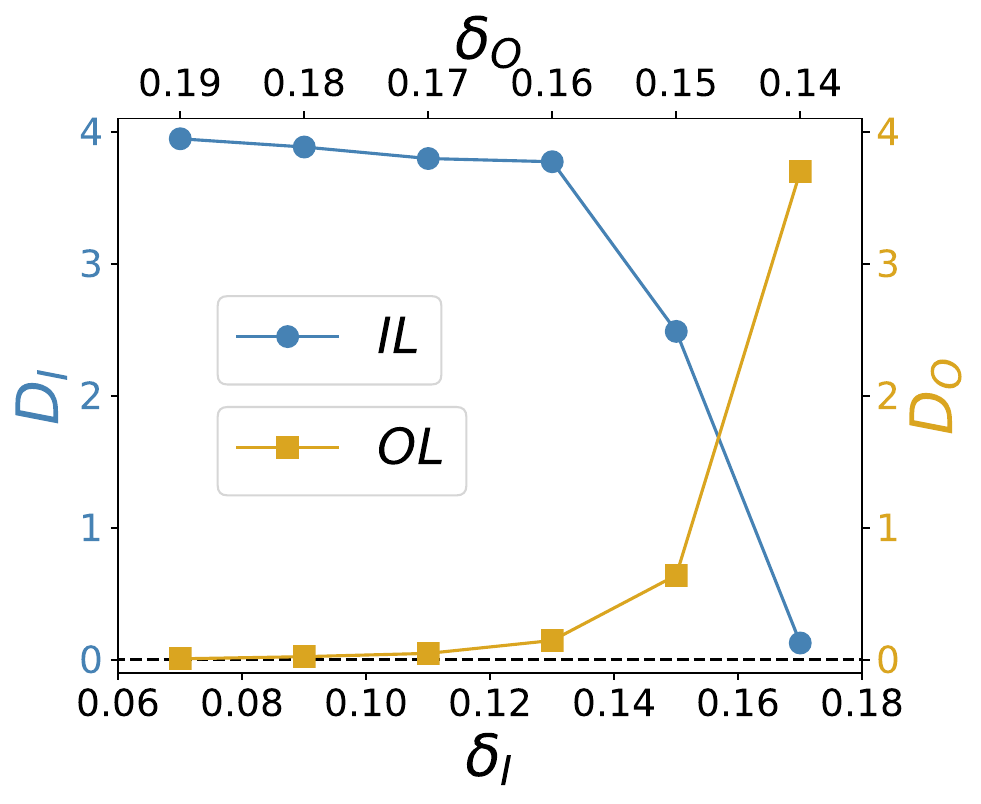,height=4.5cm,width=0.32\textwidth,clip}
\psfig{figure=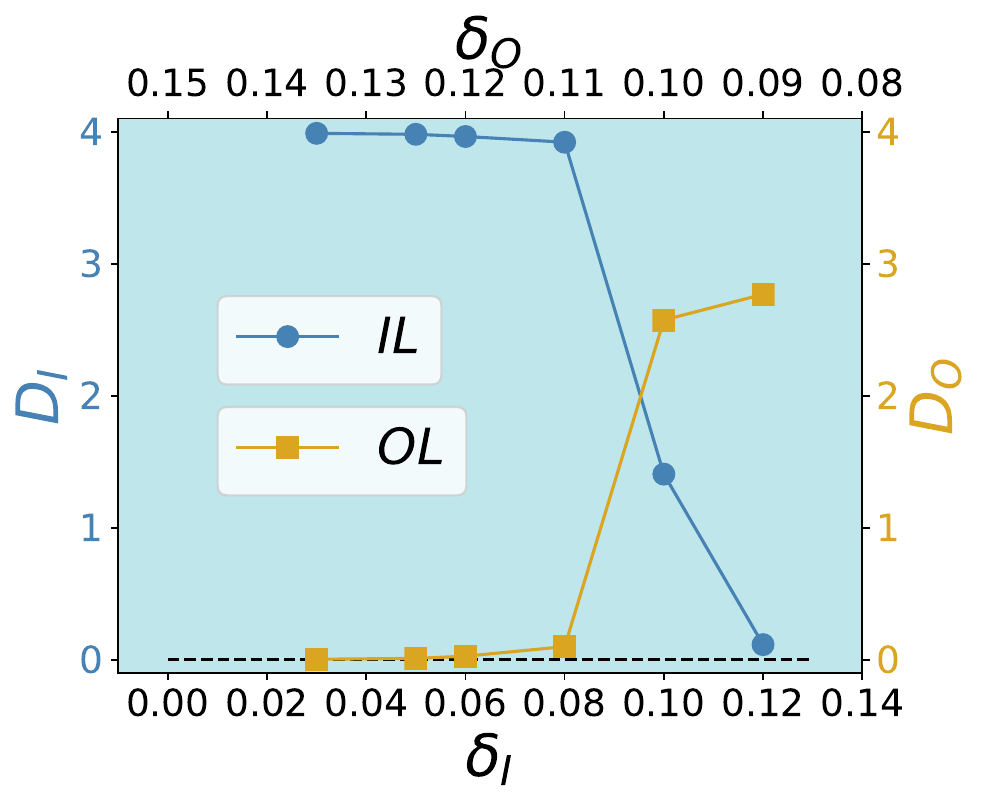,height=4.5cm,width=0.32\textwidth,clip}
\psfig{figure=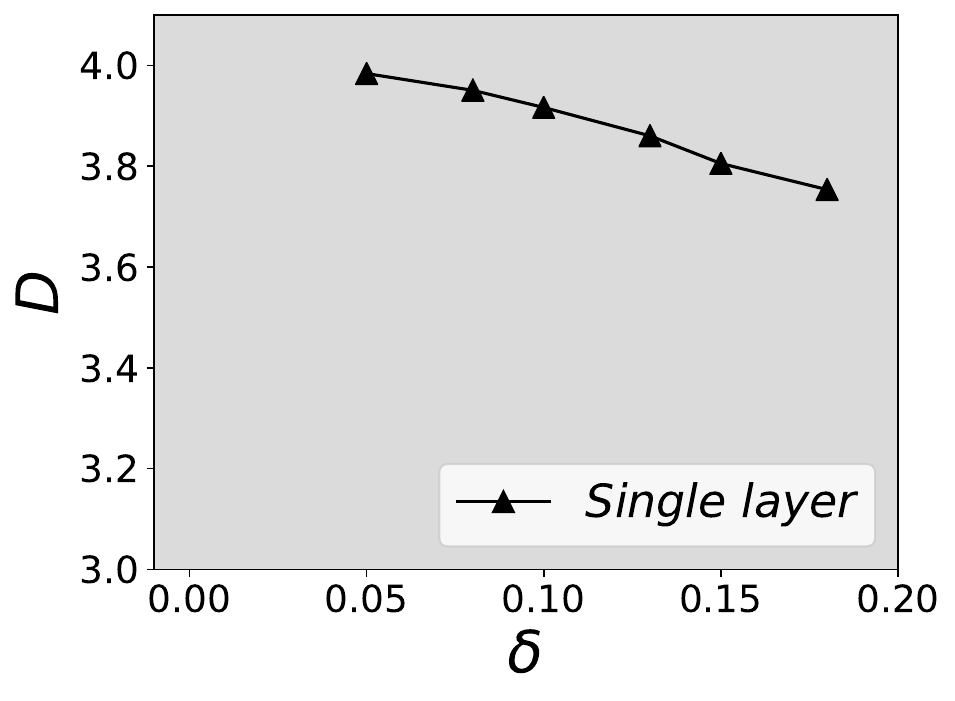,height=4cm,width=0.32\textwidth,clip}
\caption{The doping evolution of $D$ value for the IL and OL at $\delta_{avg}=0.15$ (left) and $0.1$ (middle) in trilayer model and the single layer model (right). }
\label{dvalue}
\end{figure*}

In most situations, unfortunately, we cannot directly get access to the $T_c$ limited by the QMC sign problem. As a compromise, we have to estimate the $T_c$ by extrapolating these curves to lower temperatures by assuming that they obey the logarithmic or linear temperature evolution. Fig.~\ref{Tc} provides the dependence of estimated $T_c$ on the $\delta_I$ with the upper axis indicating the corresponding $\delta_O$. As our major finding, the dominant feature lies in the optimal $T_c$ exceeding the single-layer case. For example, at $\delta_{avg}=0.1$ (orange), the highest $T_c$ occurs around $(\delta_I,\delta_O)=(0.05,0.125)$, namely the IL resides in the underdoped regime while the OL roughly at the optimal doping of single layer case (gray curve). 
Remarkably, there exists a boost of $T_c$ from the absence of SC in the single layer case to the optimal $T_c$ at $\delta_I=0.05$. 
Besides, we mention that the most recent study demonstrated the existence of a minimal doping (4\%) required for SC to appear in one of the planes~\cite{Abinitio}, which is consistent with our observation.

Moreover, both $\delta_{avg}=0.1, 0.15$ display a $T_c$ dome with varying $\delta_I$ as illustrated by the solid guiding lines in Fig.~\ref{Tc}. Note that its ending point corresponds to the case of $\delta_I=\delta_O$ with strongly suppressed $T_c$. This implies that the homogeneous hole dopings in IL and OL is detrimental to the SC, which seemingly contradicts with the previous suggestion that $T_c$ could be enhanced if all layers have balanced optimal dopings~\cite{imbalance}. This issue is possibly attributed to our limited knowledge on the optimal hole doping or the limitation of the present simplest trilayer model on accounting for the realistic cuprates, which deserves further investigation.
Furthermore, when $\delta_I> \delta_O$, the system enters into another qualitatively different regime. As discussed later in Fig.~\ref{eigenvector}, beyond this point, the OLs start to participate into the SC and the $T_c$ grows up again when the $\delta_O$ gradually decreases from overdoped regime to its optimal doping level.

To examine either IL or OL plays a dominant role in SC, Fig.~\ref{eigenvector} displays the BSE eigenvectors for three orange points corresponding to the decreasing $T_c$ segment (orange solid line) in Fig.~\ref{Tc}.
For our most interested case of $\delta_I<\delta_O$ (upper and middle rows), apparently the IL (rather than the metallic OL) plays the leading role in $d$-wave SC. As our most significant evidence, the IL with such a small hole doping $\sim0.05$ can be fully responsible for the SC of the whole system. However, the role of OLs cannot be completely ruled out. One possibility is that OLs play only the minor role as the charge reservoir. Another one follows the composite picture that the OLs provide the phase stiffness via their larger doping~\cite{Kivelson, Kivelson2}. Unfortunately, it is out of our capability to decisively differentiate these two scenarios at present.
However, when the doping levels of all layers are equal, e.g. $(\delta_I,\delta_O)=(0.1,0.1)$ as shown in the bottom row, the OLs begin to participate in the SC and its pairing strength will gradually exceed that of IL. 

To better characterize the BSE eigenvector in Fig.~\ref{eigenvector}, we calculated $D=\sum_\mathbf{K}g(\mathbf{K})\Phi_d(\mathbf{K},\pi T)$ defined as the $d$-wave projection of the eigenvector~\cite{d_value} with the $g(\mathbf{K})=\cos\mathbf{K_x} - \cos\mathbf{K_y}$, whose doping evolution is shown in Fig.~\ref{dvalue}. Clearly, at either $\delta_{avg}=0.1$ or 0.15, the $d$-wave symmetry of the IL eigenvector gradually decreases with increasing doping $\delta_I$ in contrast to $D_O$ which grows accordingly. Hence, $\delta_I=\delta_O$ is the critical point for switching the roles of IL and OL, which is closely related to the crossover from one $T_c$ dome to another in Fig~\ref{Tc}. Note that the smooth crossover around $\delta_I=\delta_O$ implies the coexistence of SC in both IL and OL.

\begin{figure}
\psfig{figure=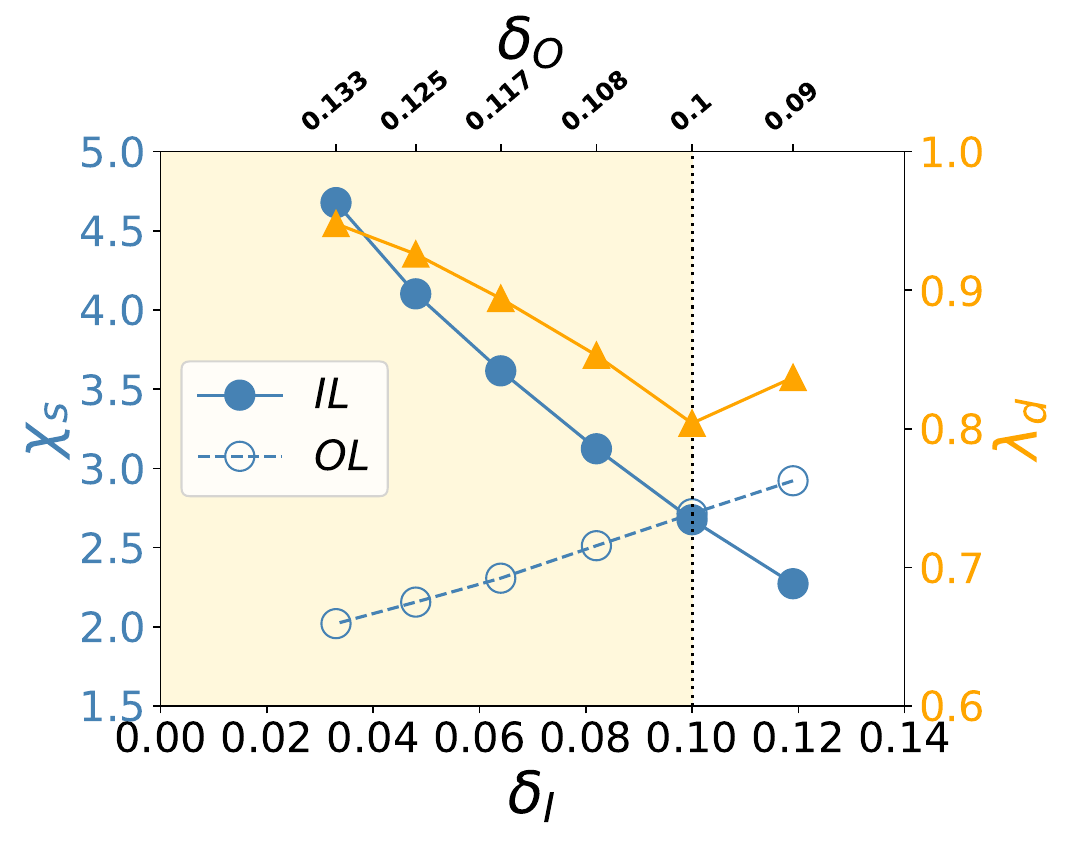,height=6.5cm,width=0.48\textwidth,clip}
\caption{Doping evolution of antiferromagnetic spin susceptibility $\chi_s(T)$ compared with the BSE eigenvalue $\lambda_d$ at our lowest simulated $T/t=0.06$.}
\label{chi}
\end{figure}

Regarding the origin of the $d$-wave SC, it is commonly believed that the antiferromagnetic (AF) spin fluctuations plays the dominant role~\cite{Maier2006,Millis2014,jm2022,Millis2022}.
To understand its different roles within IL and OL, we further examined the antiferromagnetic spin susceptibility $\chi_s$ illustrated in Fig.~\ref{chi}.  
Interestingly, in the $\delta_I<\delta_O$ regime of IL dominating SC, $\chi_s$ of IL closely follows the evolution of the orange curve of $\lambda_d$ at fixed $T/t=0.06$ (see also Fig.\ref{eigenvalue}) while the value of OLs are much smaller. This scale difference originates from the different renormalization of the electronic correlation as indicated in the self-energy in Fig.~\ref{2layers}.
This trend is turning around the other way in the regime of $\delta_I>\delta_O$, where the role of OLs in SC gradually exceeds that of IL as discussed above.
These observations strongly imply the decisive role of   AF spin fluctuations in either IL or OL dominated regimes of SC.

\section{summary and outlook}

Inspired by the highest $T_c$ in multilayer cuprates, we have investigated the trilayer Hubbard model with tunable imbalanced hole dopings in IL and OL at two characteristic average hole dopings $\delta_{avg}=0.1, 0.15$. 
By employing dynamical cluster quantum Monte Carlo calculation, our main findings as shown in Fig.~\ref{Tc} demonstrate that the imbalanced hole dopings are generically beneficial for SC. In the realistic regime of $\delta_I<\delta_O$, the IL hosts the PG behavior rather than the metallic OL. As revealed by the analysis of BSE eigenvector, we identify that the higher global $T_c$ arises from the IL itself at our lowest accessible temperature scale. 
In addition, it is worthwhile noting that the OLs can also participate in SC in the regime of $\delta_I>\delta_O$. Our investigation
also demonstrated the existence of a minimal doping in IL required for SC.

Our most remarkable observation is the boost of $T_c$ to be higher than the single layer model. One inspiring interpretation could be the capacitive coupling between the superconducting IL and metallic OL~\cite{capacitive}, which suppresses the superconducting fluctuation within IL and thereby enhance the SC in the sole IL~\cite{pengrui}. Although our model has an artificial parameter $e_I$ to tune the site energy difference between IL and OL for inducing an effective capacitive coupling, our interlayer hybridization, albeit small $t_{\perp}=0.1t$, would complicate the picture and requires more exploration. 

The role of OLs cannot be completely ruled out. One possibility is that OLs play only the minor role as the charge reservoir. Another thought follows the composite picture that the OLs provide the phase stiffness via their larger dopings~\cite{Kivelson, Kivelson2}. At present, unfortunately, it is out of our capability to decisively differentiate these two scenarios, which deserves more exploration. 
For example, it is possible that the OL can non-trivially support an emergent SC gap at the lower temperature scale closer to $T_c$.
Another attractive idea might be related to the early proposal that high $T_c$ could be realized in systems consisting of alternating layers of a bad metal with strong pairing to induce local SC and a good metal to screen the Coulomb interaction~\cite{EK1995}. 
Besides, the requirement of imbalanced IL and OL hole dopings in our study seems to contradict the previous suggestion that $T_c$ could be enhanced if all layers have optimal balanced dopings~\cite{imbalance}. 
Future possible directions also include the more challenging calculations to differentiate the roles of pairing strength and phase stiffness in IL and OL.

\section{Acknowledgment}
We acknowledge insightful discussion with Rui Peng, Yuan Li, and Yiwen Chen on the experiments of multilayer cuprates. We also thank Steven Kivelson, Dror Orgad, and Wenxin Ding for valuable comments.
This work was supported by National Natural Science Foundation of China (NSFC) Grant No. 12174278 and Priority Academic Program Development (PAPD) of Jiangsu Higher Education Institutions.

\section{Appendix}
We provide more details on the DCA methodology.
Precisely, DCA evaluates various physical observables via single- and two-particle Green functions in the thermodynamic limit via mapping the lattice problem onto a finite cluster embedded in a self-consistent mean-field bath~\cite{Hettler98,Maier05}, which is realized by the convergence between the cluster and coarse-grained (averaged over a patch of the Brillouin zone around a specific cluster momentum $\mathbf{K}$) single-particle Green's functions. 
In this manner, the short-range interactions within the cluster are treated exactly with various numerical techniques, e.g. CT-AUX in our present study; while the longer-ranged physics is approximated by a mean field hybridized with the cluster. In principle, larger cluster size systematically approaches the exact result in the thermodynamic limit in spite of its inapplicability in practice. 
The finite cluster size is essentially an approximation of the whole Brillouin zone by a discrete set of $\mathbf{K}$ points.
%so that the self-energy $\Sigma(\mathbf{K},i\omega_n)$ is a constant function within the patch around a particular $\mathbf{K}$ and a step function in the whole Brillouin zone.
%Generically, the quantum embedding methods including DCA have better minus sign problem than the finite-size QMC simulations because of the hosting mean field.
%We refer to Ref.~\cite{Maier05} for more discussions on DCA technique and its insight on the strongly correlated electronic systems.

We adopt two complementary methods to identify the pseudogap (PG) behavior. Firstly, we rely on the proxy of the desired real-frequency spectral function $A(\textbf{K},\omega=0)$=$-\frac{1}{\pi}$Im$G(\textbf{K},\omega=0)$ by the linear extrapolation of the imaginary-frequency Im$G(\textbf{K},i\omega_n$) from our DCA simulations at its two lowest Matsubara frequencies~\cite{Aw,WuWei2018}. In addition, we also calculated the spectral function $A{(\textbf{K},\omega})$ via the analytic continuation maximum entropy method~\cite{maxentropy}.

To investigate the superconducting, charge, and magnetic instability of a particular model Hamiltonian, one has to determine the structure of the interaction responsible for these channels. Essentially, the cluster two-particle Green's function 
\begin{align} %\label{e1}
  \chi_{c\sigma\sigma'}(q,K,K') &= \int^{\beta}_0 \int^{\beta}_0 \int^{\beta}_0 \int^{\beta}_0 d\tau_1 d\tau_2 d\tau_3 d\tau_4 \nonumber \\
  & \times e^{i[(\omega_n+\nu)\tau_1 -\omega_n\tau_2 +\omega_{n'}\tau_3 -(\omega_{n'}+\nu)\tau_4]} \nonumber \\
  \times \langle \mathcal{T} & c^{\dagger}_{K+q,\sigma}(\tau_1) c^{\phantom{\dagger}}_{K\sigma}(\tau_2) c^{\dagger}_{K'\sigma'}(\tau_3) c^{\phantom{\dagger}}_{K'+q,\sigma'}(\tau_4) \rangle
\end{align}
with conventional notation $K=(\mathbf{K},i\omega_n)$, $K'=(\mathbf{K'},i\omega_{n'})$, $q=(\mathbf{q},i\nu)$ and the time-ordering operator $\mathcal{T}$ can be calculated numerically via a DCA cluster solver (CT-AUX in our case).
Then the cluster two-particle irreducible vertex $\Gamma_{c\sigma\sigma'}(q,K,K')$ can be extracted through the Bethe-Salpeter equation (BSE)
\begin{align} \label{e2}
  \chi_{c\sigma\sigma'}(q,K,K') &= \chi^0_{c\sigma\sigma'}(q,K,K') + \chi^0_{c\sigma\sigma''}(q,K,K'') \nonumber \\
  & \times \Gamma_{c\sigma''\sigma'''}(q,K'',K''') \chi_{c\sigma'''\sigma'}(q,K''',K')
\end{align}
where $\chi^0_{c\sigma\sigma'}(q,K,K')$ is the non-interacting two-particle Green's function constructed from the product of a pair of fully dressed single-particle Green's functions. The usual convention that the summation is to be made for repeated indices is adopted. 

Note that the above formalism Eqs.~(1-2) has their counterparts for the corresponding lattice quantities, whose numerical calculations are, however, impractical due to their continuous nature. Therefore, one of the key DCA assumptions is that the cluster two-particle irreducible vertex $\Gamma_c$ is used as the approximation of the desired lattice two-particle irreducible vertex $\Gamma$.

The two-particle irreducible vertex and associated BSE Eq.~\eqref{e2} can be classified according to the superconducting, charge, and magnetic channels. In this work, we are mostly interested in the particle-particle superconducting channel for the zero center-of-mass and energy.
To this aim, the superconductivity can be quantitatively displayed by the leading eigenvalues of the BSE in the particle-particle channel in the eigen-equation form~\cite{Maier2006,scalapino2007numerical}
\begin{align} \label{BSE}
    -\frac{T}{N_c}\sum_{K'}
	\Gamma^{pp}(K,K')
	\bar{\chi}_0^{pp}(K')\phi_\alpha(K') =\lambda_\alpha(T) \phi_\alpha(K)
\end{align}
where $\Gamma^{pp}(K,K')$ denotes the lattice irreducible particle-particle vertex of the effective cluster problem with combining the cluster momenta $\textbf K$ and Matsubara frequencies $\omega_n=(2n+1)\pi T$ as $K=(\mathbf{K}, i\omega_n)$. 
The coarse-grained bare particle-particle susceptibility
\begin{align}\label{eq:chipp}
	\bar{\chi}^{pp}_0(K) = \frac{N_c}{N}\sum_{k'}G(K+k')G(-K-k')
\end{align}
is obtained via the dressed single-particle Green's function $G(k)\equiv G({\textbf k},i\omega_n) =
[i\omega_n+\mu-\varepsilon_{\textbf k}-\Sigma({\textbf K},i\omega_n)]^{-1}$, where $\mathbf{k}$ belongs to the DCA patch surrounding the cluster momentum $\mathbf{K}$ with $\mu$ the chemical potential, $\varepsilon_{\textbf k}=-2t(\cos k_x+\cos k_y)-4t'\cos k_x\cos k_y$ the
dispersion relation, and $\Sigma({\textbf K},i\omega_n)$ the cluster self-energy. In our simulations, we chose 24 or more discrete points for both the positive and negative fermionic Matsubara frequency $\omega_n=(2n+1)\pi T$ mesh for measuring the four-point quantities like two-particle Green functions and irreducible vertices. Therefore, the BSE Eq.~\eqref{BSE} reduces to an eigenvalue problem of a matrix of size $(48N_c)\times (48N_c)$.

The calculated leading eigenvalue $\lambda_\alpha(T)$ reflects the pairing tendency with symmetry $\alpha$ in the normal state. Simultaneously, the associated eigenvector $\phi_\alpha(K)$ can be viewed as the normal state analog of the SC gap function~\cite{Maier2006,scalapino2007numerical}.

In addition to the particle-particle channel discussed above, we are also interested in particle-hole vertex $\Gamma^{ph}$ in magnetic channels since the dominant contribution to the $d$-wave pairing interaction has been shown to arise from the spin-one ($S=1$) particle-hole exchange.
Therefore, we extracted the irreducible particle-hole vertex $\Gamma^{ph}$ in the magnetic channel of the effective cluster problem and thereby we have the BSE in the eigen-equation form similar to Eq.~\eqref{BSE} but with coarse-grained bare particle-hole susceptibility
\begin{align}\label{eq:chiph}
	\bar{\chi}^{ph}_{0}(q,K,K') = \delta_{KK'} \frac{N_c}{N} \sum_{k'}  G(K+k') G(K+k'+q)
\end{align}
The corresponding eigenvalues for the particle-hole channels reflect the tendency of magnetic instability.

In this work, we are only interested in the case of zero frequency transfer ($i\nu=0$) similar to the particle-particle superconducting channel Eq.~\eqref{BSE}. Besides, we focus on the $(\pi,\pi)$ momentum transfer, namely the antiferromagnetic lattice susceptibilities. This can be obtained by the coarse-grained two-particle Green function $\bar{\chi}^{ph}(q,K,K')$ (instead of cluster quantities that result in cluster susceptibilities), which is in turn calculated via the coarse-grained BSE transformed from Eq.~\eqref{e2} as
\begin{align}\label{}
	[\bar{\chi}^{ph}(q,K,K')]^{-1} = [\bar{\chi}^{ph}_{0}(q,K,K')]^{-1} - \Gamma^{ph}(q,K,K')
\end{align}
Then our interested magnetic lattice susceptibilities $\chi_s(q,T)$ can be deduced as
\begin{align}\label{chics}
	\chi_s(q,T) = \frac{T^2}{N_c^2} \sum_{K,K'} \bar{\chi}^{ph}(q,K,K')
\end{align}

\bibliography{main.bib}

\end{document}